\documentclass[aps,prb,twocolumn]{revtex4}
\usepackage{epsfig}

\newcommand{\fig}[1]{Fig.~{\ref{fig:#1}}}

\begin{document}

\title{A selected history of expectation bias in physics}

\author{Monwhea Jeng}

\email{mjeng@siue.edu}

\affiliation{
Physics Department, Box 1654,
Southern Illinois University Edwardsville,
Edwardsville, IL, 62025}

\begin{abstract}
The beliefs of physicists can bias their results towards 
their expectations in a number of ways.  We survey a variety
of historical cases of expectation bias in
observations, experiments and 
calculations.
\end{abstract}

\maketitle


\smallskip

{\narrower\noindent\sl
It is a capital mistake to theorise before one has data.
Insensibly one begins to twist facts to suit theories,
instead of theories to suit facts.\\

\hspace{2.0in} Sherlock Holmes}~\cite{Holmes}

\smallskip

{\narrower\noindent\sl
But are we sure of our observational facts? Scientific
men are rather fond of saying pontifically that one ought
to be quite sure of one's observational facts before embarking
on theory. Fortunately those who give this advice do not
practice what they preach. Observation and theory get on
best when they are mixed together, both helping one another
in the pursuit of truth. It is a good rule to not put
overmuch confidence in a theory until it has been confirmed
by observation. I hope I shall not shock the experimental
physicists too much if I add that it is also a good rule 
not to put overmuch confidence in the observational results 
that are put forward until they have been confirmed by theory.\\

\hspace{1.55in} Sir Arthur S. Eddington}~\cite{Eddington}

\smallskip


\section{Introduction}
\label{sec:intro}

Under an idealized view of science, the theoretical 
beliefs of experimenters should have no effect on what results 
are obtained.  Theoretical beliefs can influence what experiments 
are done, and how they are interpreted and recieved, but 
the actual results are facts determined purely by 
nature. But in practice, one can worry that even honest and
careful experimenters might still somehow transfer their
biases to their end results.

Sherlock Holmes warns against this by forbidding any
theorizing in advance of the experiment. But it is difficult
to see how this is possible, let along desirable. And
when we remember that Sir Arthur Conan Doyle, the author 
of {\it Sherlock Holmes}, uncritically accepted, and indeed 
actively promoted, the ``data'' that fairies had been seen 
and photographed in Cottingley, England~\cite{fairies}, we 
begin to think that perhaps not all ``experimental facts'' 
should be accepted and treated equally. 
Eddington's characterization of the 
complex interplay between theory and experiment seems to 
come closer to the truth.  Perhaps the facts from experiments 
and observations are not always simple ``facts,'' to be 
accepted at face value.

Expectation bias is well-known, and widely discussed
in the ``softer'' sciences, that study humans,
such as psychology and medicine. But what role does it
play in ``hard'' sciences, such as physics?
The impression one gets from the history in physics
textbooks is that it plays virtually no role, but this may
be misleading. 
Because textbooks have as their primary goal to
teach physics, extended discussions of how this 
physics was arrived at can seem like pointless and
potentially
confusing diversions; the emphasis is on how careful
and correct reasoning leads to correct results,
even if that reasoning is retrospective, and ahistorical.
Convoluted reasoning that was actually followed is replaced
with the clearer
reasoning that ``should'' have been followed. 
Many readers will doubtlessly be surprised to discover
that Planck was not led to quantization in an attempt
to fix an ``ultraviolet catastrophe'' discovered by
Rayleigh~\cite{History.Brush,History.Whitaker}, 
that R\o mer never calculated a value for
the speed of light~\cite{Romer}, and
that Einstein was not primarily motivated by the
Michelson-Morley experiment in his invention of special
relativity~\cite{relativity}.
These, and
other examples, are discussed in the
reviews by Brush and Whitaker~\cite{History.Brush,
History.Whitaker}.
While this systematic bias in textbooks is understandable,
it unfortunately tends to eliminate cases of expectation 
bias.  Here, we consider some historical cases where beliefs 
of physicists have influenced their results. 


\section{Observations}
\label{sec:perception}

When human perception is at its limits, even basic observations 
are unreliable, and may be biased towards what the observer 
expects to see. Van Helden surveyed the history of early telescopic 
observations of Saturn~\cite{Saturn}. While most observers 
today, even with very low-resolution pictures, see a 
planet with rings, early observers echoed Galileo's first 
observations, and saw and reported on a planet with 
two giant moons, one on each side. Van Helden's analysis
convincingly shows that the inaccurate drawings of Saturn
came not from poor telescopes, but from the influence of 
Galileo's first reports on later observers. Van Helden
traces the evolution of observations from a ``two moon'' 
phase to a ``ring'' phase.

Another example can be seen in the N-rays of the
early 1900's~\cite{Nrays}.
From 1903 to 1906, roughly 300 papers on N-rays were written
by 100 scientists and doctors throughout France,
describing in detail the properties of N-rays, as deduced by
the faint lines they produced on white surfaces in dark
rooms.  However, many researchers were unable to see these 
lines.  While there were early suspicions,
it was not until several years had passed
that the scientific 
community came to decide that the positive results arose
from experimenters convincing themselves that they 
saw faint lines and flashes that were not really there.
In fact, anyone staring at
a white surface under such conditions
long enough will start to see dark spots
and lines, and the judgment as to what constituted a
flash turned out to be rather subjective.
While this case has been used as an example of
``pathological science,''~\cite{pathological}
it is not clear how to distinguish
pathological science from normal science in error.

As Brewer points out, in almost all cases like these,
the observations required were at the limits of human 
perception, and it is in these sorts of cases where we can 
expect subconscious biases to seep in~\cite{Brewer}.  When 
the perceptual data are unambiguous, these sorts of problems 
will generally not occur; no experimenter is likely to report 
that apples fall up. However, it is not always clear what 
constitutes ``unambiguous'' perceptual data.  Bisson and 
Dennen noted that while Newton's experiments with color 
used a prism under conditions where spectral lines 
should have been clearly visible, Newton never reported 
seeing such lines~\cite{NewtonColor1}.  Boring argued that 
Newton's presuppositions essentially prevented him from 
seeing these spectral lines~\cite{NewtonColor2}.  Similarly, 
it is striking that there are no European records of the 
1054 supernova. This supernova was visible to the naked 
eye by daylight for several weeks, and at night for over 
a year, and appears in Chinese and Japanese records.  It 
has often argued that Europeans failed to record and discuss 
the supernova because of ideological committments to an 
unchanging universe~\cite{supernova}. These arguments are 
necessarily speculative, since it is difficult to draw firm 
conclusions from the absence of records, but the 
lack of reports of
such a dramatic event is rather striking.  

But one does not need to rely on historical cases to 
see this observational bias in practice.
Physics Education Research has shown that students in
physics classes do not always see what they are
``supposed to'' in demonstrations. Instead, students
sometimes see what they expect to see, even when the
demonstration is supposed to provide a dramatic illustration
of a counterintuitive result. 
Redish describes how his students ``saw'' a marble
leaving a curved track continue to curve, instead of seeing
the straight line motion that they were supposed to
observe~\cite{Redish}.
Gunstone and White reported similar results 
when studying
student observations of experiments on gravity;
what the students saw was highly correlated
with what they expected to see~\cite{PER.gravity}.

However, even when the perceptual data is 
completely clear, other errors 
can occur. For example, when data is recorded by hand, it 
may simply be misrecorded. It might be thought that these 
errors, while regrettable, would have no overall effect 
in any direction. 
However, in 1978 Rosenthal reviewed 21 psychology studies
that allowed analysis of errors in recording. Subject
responses recorded by human observers were compared to those
recorded by mechanical devices. Roughly 140,000
observations, by over 300 observers, were looked at.
Rosenthal found that about two-thirds of the recording
errors favored the hypothesis of the observer, significantly
greater than the 50\% that would be expected if the
observers were unbiased~\cite{RecordingError}.  
Psychology differs substantially from physics, and has many
more ways in which experimenter beliefs can influence
their results~\cite{Psychology}. However, simple recording
errors should presumably play the same role in
physics experiments where individual data points can 
immediately be seen as favoring or disfavoring a hypothesis.


\section{Calculations}
\label{sec:miscalculations}

However, in many physics experiments, data can only be seen
as favoring or disfavoring a hypothesis after lengthy calculations. 
This would seem to prevent recording errors from being biased 
in any particular direction.  However, this will not save 
us from experimenter biases, because errors then can occur 
in calculations, and just as with the recording errors,
will tend to favor experimenter predispositions. 

The first experiment to observe and quantitatively 
measure the pressure due to light, by Nichols and Hull, 
found agreement with Maxwell's theory to within 1\%.
However, Bell and Green later reanalyzed the data 
from these experiments, and found several mathematical errors: 
Nichols and Hull had used an incorrect value for the mechanical 
equivalent of heat, had taken some logarithms to base ten 
instead of base $e$, and made several mistakes over units 
and conversion factors. Once these mistakes were corrected, 
the results deviated from Maxwell's theory by
10\%, which was still a success, but not nearly as 
good as the original 1\% agreement 
reported~\cite{LightPressure}.
It seems reasonable to think that Nichols and Hull, 
seeing such good agreement with Maxwell's theory,
had been happy to publish, but that if the mathematical
errors had been in the opposite direction,
they would have checked over their calculations more
carefully.

More subtle errors occurred in the searches for free quarks by
Fairbank and his collaborators~\cite{blind}. They measured
the charges on several niobium spheres, obtaining
$(-0.343\pm 0.011)e$, $(0.001\pm 0.033)e$, 
and $(+0.328\pm 0.007)e$, in excellent agreement
with the fractional charges of $\pm e/3$ expected for 
quarks.  However, other experimenters failed to find evidence 
of free quarks. The
calculations needed to turn Fairbank's raw data into charges
were quite complex, and it was suggested that the Fairbank 
group 
might have been unconsciously biased by their expectations 
when doing them. The 
Fairbank group thus did a ``blind'' 
analysis.  To do this, they added a random offset 
to their original data, and recalculated all charges
without knowing the value of this offset. Only when 
all calculations were completed was the offset
revealed and removed.
When this was done, the calculated 
charges were instead
$(+0.189\pm 0.02)e$, and $(+0.253\pm 0.02)e$,
which did not agree with the quark model, or 
indeed, any major theoretical model.
This case, and other cases of blind analysis
in physics,
can be found in the book by Franklin~\cite{blind}.

The effects of expectation bias can be particularly strong
in cases where the physicists have a great deal of freedom
in their analysis, such as when ``eyeballing'' fits.
In the 1920's, Millikan and Cameron ran experiments on cosmic 
rays, to test Millikan's theory of the ``birth cry of atoms,'' 
which postulated that heavy elements were formed in the 
space between the stars~\cite{Galison}.  Not only was this 
theory false, but they analyzed their cosmic ray data with 
absorption formulae from Dirac that were subsequently discovered to 
be in error.  Nevertheless, Millikan and Cameron were able 
to obtain excellent agreement between their experimental 
results, and the ``birth cry of atoms'' theory! Their results 
gave quantitatively precise agreement for the production 
of oxygen, nitrogen, helium, and silicon, and they obtained 
rough agreement for the production of iron~\cite{MillikanBirth}.  
They obtained these results by measuring ionization from
cosmic rays as a function of depth, and then fitting by eye
this absorption curve as a sum of three exponentials. But
such a fit gave them excessive freedom, so that their final
results for the coefficients and exponents of these three
exponentials, were essentially arbitrary.

Calculations influenced by the desire to
obtain certain results even date back to Newton's
{\it Principia}~\cite{NewtonPrincipia}.  
Newton's calculation of the speed of sound in air agreed
with experiment to within 0.1\%.
This calculation of the speed of sound
was based on the assumption that the the air underwent isothermal 
compressions and expansions.  However, correct treatment 
of sound in air requires treating the compressions and expansions 
as adiabatic, so Newton {\it should} have obtained a 
speed of sound 15\% too low. 
It is often said that Newton did obtain a speed of
sound 15\% too low; see, for 
example Ref.~[\cite{SoundText}]. 
But he did not: he obtained results that agreed
perfectly with the experiments at the time!
He did this by applying a number of
poorly-motivated ``corrections'' to his calculations,
based on quantities he could not possibly have known (and
did not
know), such as the finite volume of space taken up by air
molecules, and the effects of the water vapor in the
air. He thus calculated corrections to the isothermal result
that magically achieved a 0.1\% agreement
with the most recent experimental results
available to him.

An interesting postscript appears in the later correction 
to Newton's calculation of the speed of
sound~\cite{Laplace.sound.1,Laplace.sound.2,Caloric}.  
Other scientists 
recognized Newton's fudging for what it was, and the discrepancy 
between the theoretical and experimental results for the 
speed of sound remained an outstanding problem for over 
100 years. In the early 1800's, Laplace, armed with
recent experimental results on specific 
heat capacities of gasses by Delaroche and Berard, gave 
an adiabatic treatment of sound waves, deriving a speed 
of sound differing only 2.5\% from experiment.
But not only did Laplace use the
now-discredited caloric 
theory in his determination of $\gamma$,
but we now know that the experimental
results of Delaroche and Berard are 12\% off
the corrent values~\cite{CaloricNote}!
The two 
errors ended up canceling, giving a spurious agreement.
Here the agreement seems to be fortuitous, 
rather than the result of bias.


\section{The Bandwagon Effect}
\label{sec:bandwagon}

In other cases, experimenter bias is clear, but it is more 
difficult to identify precisely where the bias crept in. 
When, in 1915, Einstein 
and de Haas performed the first experiments measuring the 
gyromagnetic ratio of the electron, they expected, on the 
basis of their models, to find a $g$-factor of $g=1$.
We know today that the correct value is roughly 2,
yet Einstein and de Haas obtained $1.02\pm
0.10$~\cite{Galison}!
Around the same time, the American scientist
Barnett independently did two sets of experiments on
magnetism, and found
$g=2.0$, and $g=2.3$. But after hearing of the results 
of Einstein and de Haas, Barnett repeated his experiments,
and then
reported 
that $g$ was between 1.1 and 1.4, stating that 1.0 
was within his error bars.  Subsequent experiments by other 
groups did obtain values closer to 2. However, de Haas, 
in three more experiments in 1915 and 1923, continued to
obtain values 
significantly below 2 (1.2, 1.11, and 1.55).  It seems clear 
that Einstein, de Haas, and Barnett,
were somehow influenced by their expectations.
A very enjoyable history of this episode, as well as 
the already-mentioned Millikan ``birth cry of the atoms'' 
episode, can be found in the book 
by Galison~\cite{Galison}.
Galison speculates that Einstein and de Haas may have 
unintentionally corrected for systematic errors in a
biased fashion, fixing systematic errors that led to 
higher values of $g$, but leaving alone ones that led to
lower values of $g$.

A similar problem can be seen in the history graphs
produced by the Particle Data Group's
{\it Review of Particle Properties}.
This group produces a report of all physical properties
every two years, and their history graphs show
reports of certain particle properties 
as a function of the year of their report.
Ideally, the reported values should be randomly 
scattered about the constant correct value. However,
some of the graphs show distinct trends
with time~\cite{ParticleDataGroup}.
The 1980 Particle Data Group explained 
this as follows~\cite{PDG.Explained}:

\smallskip
{\narrower
\noindent We show these figures not only to demonstrate 
that there is not much change in these averages in the usual
case, but also to show that there exist cases with
relatively large changes. There is a psychological danger in
preparing tables of ``right'' answers. The old joke about
the experimenter who fights the systematics until he or she
get the ``right'' answer (read ``agrees with previous
experiments''), and then publishes, contains a germ of
truth (presumably, those who compile and average
experimental results are also not immune to this disease).
A result can disagree with the average of all previous
experiments by five standard deviations and still be right!
Hence, perhaps it is of value to show that large changes
can (and do) sometimes occur.

}
\smallskip

\begin{figure}[tb]
\epsfig{figure=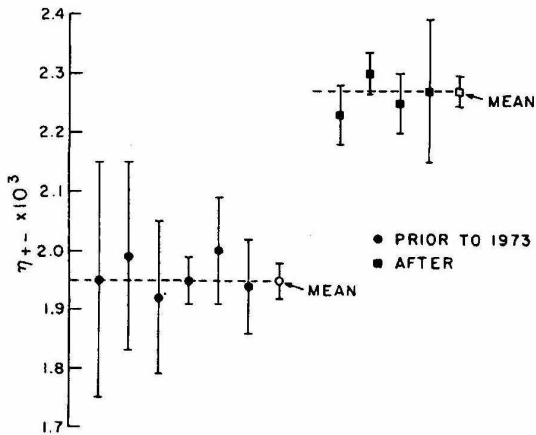,width=3.0in}
\caption{Measurements of $\mid\eta_{+-}\mid$ in order of 
their year of publication.  Reprinted with permission from 
A. Franklin, ``Forging, cooking, trimming, and riding on 
the bandwagon,'' Am. J. Phys.  {\bf 52}, 786-793 (1984), 
copyright 1984, American Association of Physics Teachers.}
\label{fig:CP}
\end{figure}

\noindent Franklin terms  
these time-dependent shifts and trends ``bandwagon 
effects,'' and 
discusses a particularly dramatic case with 
$\mid\eta_{+-}\mid$, 
the parameter that measures CP 
violation~\cite{Bandwagon}. 
As seen in \fig{CP}, measurements of 
$\mid\eta_{+-}\mid$ before 1973 are systematically 
different from those after 1973.

\begin{figure}[tb]
\epsfig{figure=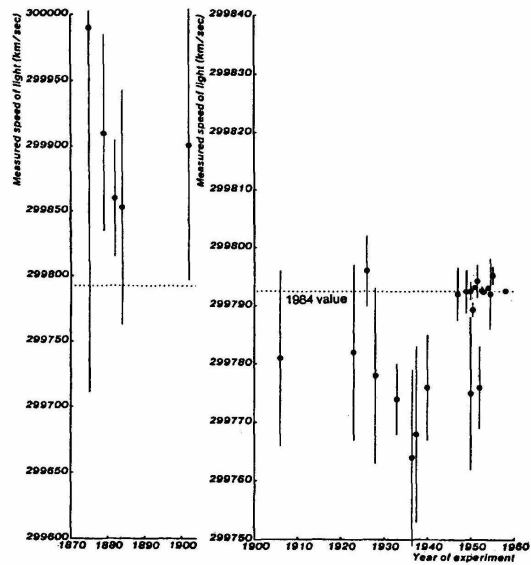,width=3.0in}
\caption{Measurements of the speed of light, as a function 
of the year of the experiment.  Note that the vertical scale 
is changed at 1900, to make the errors bars visible.
Reprinted with permission from M. Henrion and B. Fischhoff,
``Assessing uncertainty in physical constants,'' Am. J. 
Phys. {\bf 54}, 791-798 (1986), copyright 1986, American 
Association of Physics Teachers.}
\label{fig:SpeedLight}
\end{figure}

Henrion and Fischhoff
graphed experimental reports on the speed of light,
as a function of the year of the experiment.
Their graph (\fig{SpeedLight})
shows that experimental results
tend to cluster
around a certain value
for many years, and then suddenly jump to cluster
around a new value, often many error bars from the
previously accepted value~\cite{HenrionFischhoff}.
Two such jumps occur for the speed of light.
Again, the bias is clearly present in the graph, but
the exact mechanism by which it occured is
harder to discern.

\section{Results that are too good to be true}
\label{sec:too.good}

Newton's results on the speed of sound, and Millikan and 
Cameron's results on cosmic rays, are in retrospect too good
to be true,
because they were working with 
deficient or incorrect theories. But even results that
agree with correct theories can be too good to be true.

Henrion and Fischhoff looked at the 
reported values for a number of other fundamental constants, 
and with one exception, found that error bars had been systematically
underestimated---results were more spread about the correct 
value than they should have been, based on the reported 
error bars~\cite{HenrionFischhoff}.  Underestimating error 
bars is not necessarily a sign of experimenter bias.  However,
their one exception is. Unlike the fundamental constants 
such as $\hbar$ or $c$, for which there are 
no theoretical calculations to tell experimenters what to 
expect, experiments measuring the ratio of the absolute 
ohm to the ohm as maintained by the National Bureau of
standards ($\Omega_{ABS}/\Omega_{NBS}$) ``should'' obtain 
one, if all calibrations are correct.  Henrion and Fischhoff 
found that the experimental results for $\Omega_{ABS}/\Omega_{NBS}$ 
were too closely clustered around one---given the error 
bars reported, the chance was only 1 in 200 that the results 
would be so clustered together.  This indicates that experimenters, 
knowing that they should obtain one, somehow biased their 
measurements or calculations to get ``the right answer.'' 

We depart briefly from physics to consider
a famous case from biology.
Fisher's reexamination of Mendel's results
in his famous experiments on heredity in peas,
showed that Mendel's results were simply too good to be 
true.
Many of his results show ratios that almost exactly
match the theoretical values, with deviations much
smaller than the expected statistical fluctuations.
Fisher showed that only 3 experiments in 100,000
would obtain results as close to theory as
Mendel's~\cite{MendelFisher,MendelNote}.
This has often been taken
as proof of fraud, either on the part of Mendel, or
an assistant, but no evidence of fraud has been found,
other than the fact that the data is too good to be true.
As we have seen, such biases can
arise in a number of ways without intentional fraud.
For example, Root-Bernstein has shown that the results can 
be explained if Mendel was unconsciously biased in cases 
where the assignation of pea traits was 
ambiguous~\cite{MendelRoot}.
Olby has pointed out that the results could be explained
if Mendel kept track of the ratios as he collected 
data, and stopped recording when the ratios 
were what he expected~\cite{MendelOlby,MendelOlbyNote}.


\section{Other cases}
\label{sec:misc}

Experimenter biases can also arise from ambiguities
in when to discard data. If equipment was clearly 
malfunctioning during a run, data can no longer 
be used. But it will sometimes be unclear 
whether data should be discarded, and in ambiguous cases, 
whether the data agrees with expectations may 
unconsciously (or consciously) tip 
the experimenter in one direction or another.  One of the 
most (in)famous cases is that of Millikan's oil drop 
experiments, where he both demonstrated the quantization of 
charge, and measured this quantized 
charge~\cite{MillikanOilDrop}.
In his paper he claimed ``{\it It is to be remarked, too,
that this is not a selected group of drops but represents
all of the drops experimented on during 60 consecutive
days}. . .''~\cite{MillikanOriginal}.
But later inspection of his lab 
notebook shows that he threw out results for a number of 
oil drops, in many cases after calculating their charge.  
For some oil drops the equipment was clearly malfunctioning.
For others, the data failed certain consistency checks. And 
for a few drops, Millikan simply discarded 
them because the calulated charges were too far off from 
what he believed to be the correct value.  The 
exclusion of the latter data points appears to cross the 
line into fraud, but the reasons used to discard data form 
a continuum from obviously reasonable to fraudulent, and 
illustrate the difficulty in determining when it is appropriate 
to discard data.
(Fraud is, of course, also a way in which experimenter
beliefs can influence their results, but discussion of
such cases is beyond the scope of this paper.)

Here we have only discussed the ways in which theoretical
beliefs can affect the actual results reported.
Once the results are reported, theoretical
beliefs can also affect how the results are interpreted,
and whether they are believed and/or published. We will not
discuss such cases here, except to note
that it is not always easy to separate the cases where
theoretical beliefs have affected the experiment,
and when they have merely affected its interpretation.
The difficulty of this distinction is well-illustrated
by an episode in the development of the kinetic theory
of gasses.

In 1859 the scientific understanding of heat was still
uncertain, and the atomic picture for materials was
still controversial. Maxwell considered his 
now-famous model of a gas as a collection of billiard 
balls (atoms) undergoing elastic collisions, and came up with
a startling result---he showed that with this model,
the viscosity of a gas would be independent of its 
density~\cite{MaxwellKinetic}.
This was surprising, since 
he and his colleagues, naturally,
expected the viscosity to go to zero as the density
went to zero.
Maxwell asked Stokes about this, and Stokes assured him
that experiments by Sabine on the damping of pendulums in air
had demonstrated that the viscosity of air
went to zero at low densities---apparently,
a strong disconfirmation of the billiard ball model!
Nevertheless, Maxwell continued his investigations,
and eventually did experiments showing that gas viscosity
was constant over a wide range of densities.
With this triumph of the billiard ball model, Stokes went
back and reanalyzed Sabine's experiment. Stokes came to
realize that, in fact, his previous analysis of Sabine's
experiment has implicitly {\it assumed} that the viscosity
was proportional to density at low densities, thus
assuming the result he claimed to be proving.
It could be argued that this is not
a case where theory affected reported experimental results, but
merely one where it affected the way in which results were 
interpreted; but the line is blurry, for Stokes and Maxwell 
both apparently believed that ``the viscosity goes to zero 
as the density goes to zero,'' was the experimental result.

We have seen that there are a number of ways in
which the theoretical expectations of
physicists can influence their results.
This bias is certainly less prominent than in fields such
as psychology and medicine, but it is clearly present,
and important if we want to understand how physics works in
practice. Finally, if the reader knows of any cases of
expectation bias, either historical or recent, in their own
fields, I'd love to hear about them!

\acknowledgments{This research was supported in part by a 
Southern Illinois University of Edwardsville Summer
Research Fellowship. I would like to thank Robert G. Wolf
for useful discussions.}



\begin{thebibliography}{99}
\bibitem{CaloricNote} Laplace's derivation of a factor of
$\sqrt{\gamma}=\sqrt{c_p/c_v}$ in the speed of sound was
valid despite his use of caloric theory, but his method of
calculating the value of $\gamma$ from the experimental
results of Delaroche and Berard was not.
\bibitem{MendelNote} A number of authors have also taken
issue with Fisher's statistical arguments, although few
find enough problems with Fisher's analysis to conclude
that Mendel's data are unproblematic.
\bibitem{MendelOlbyNote} Root-Bernstein has pointed out a 
logical flaw in Olby's evidence that such optional 
stopping occured, but is also unable to rule it out
as a possible mechanism~\cite{MendelRoot}.
\end{thebibliography}
\end{document}